\documentclass[conference]{IEEEtran}

\usepackage{cite}
\usepackage{graphicx}
\usepackage{amsmath,amssymb}
\usepackage{subfigure}

\usepackage{url}
\usepackage{bm}
\usepackage{amssymb}
\usepackage{comment}
\usepackage{color}
\usepackage{geometry}	
\usepackage{algorithm}
\usepackage{algorithmic}


\usepackage{setspace}
\let\Algorithm\algorithm
\renewcommand\algorithm[1][]{\Algorithm[#1]\setstretch{1}}

\IEEEoverridecommandlockouts

\usepackage[acronym,shortcuts]{glossaries}
\newacronym{2D}{2D}{two-dimensional}
\newacronym{3D}{3D}{three-dimensional}
\newacronym{ADoA}{ADoA}{angle difference of arrival}
\newacronym{AoA}{AoA}{angle of arrival}
\newacronym{GPS}{GPS}{global positioning system}
\newacronym{IoT}{IoT}{Internet of Things}
\newacronym{MDS}{MDS}{multidimensional scaling}
\newacronym{MSE}{MSE}{mean square error}
\newacronym{LAN}{LAN}{local are network}
\newacronym{LOS}{LOS}{line-of-sight}
\newacronym{PDF}{PDF}{probability density function}
\newacronym{RSSI}{RSSI}{received signal strength indicator}
\newacronym{SMDS}{SMDS}{super multidimensional scaling}
\newacronym{CD-SMDS}{CD-SMDS}{complex-domain SMDS}
\newacronym{QD-SMDS}{QD-SMDS}{quaternion-domain super multidimensional scaling}
\newacronym{SVD}{SVD}{singular value decomposition}
\newacronym{QSVD}{QSVD}{Quaternion singular value decomposition}
\newacronym{ToA}{ToA}{time of arrival}
\newacronym{TDoA}{TDoA}{time difference of arrival}
\newacronym{AN}{AN}{anchor node}
\newacronym{TN}{TN}{target node}
\newacronym{GEK}{GEK}{Gram edge kernel}

\begin{document}
%

\title{Quaternion Domain Super MDS for 3D Localization}

\author{\IEEEauthorblockN{Keigo Masuoka$\,^*$, Takumi Takahashi$\,^*$, Giuseppe Abreu$\,^\dagger$, and Hideki Ochiai$\,^*$}
\IEEEauthorblockA{
$\,^*$ Graduate School of Engineering, Osaka University, 2-1 Yamada-oka, Suita, 565-0871, Japan\\ 
$\,^\dagger$ School of Computer Science and Engineering, Constructor University, Campus Ring 1, 28759 Bremen, Germany\\
    Email: 
    $\,^*$\{masuoka@wcs., takahashi@, ochiai@\}comm.eng.osaka-u.ac.jp, $\,^\dagger$gabreu@constructor.university\\[-2ex]
}}

\maketitle
\begin{abstract}
%
%
We propose a novel low-complexity \ac{3D} localization algorithm for wireless sensor networks, termed \ac{QD-SMDS}.
This algorithm reformulates the conventional \acs{SMDS}, which was originally developed in the real domain, into the quaternion domain.
By representing \ac{3D} coordinates as quaternions, the method enables the construction of a rank-1 \ac{GEK} matrix that integrates both relative distance and angular (phase) information between nodes, maximizing the noise reduction effect achieved through low-rank truncation via \ac{SVD}.
The simulation results indicate that the proposed method demonstrates a notable enhancement in localization accuracy relative to the conventional \acs{SMDS} algorithm, particularly in scenarios characterized by substantial measurement errors.

%
%
%
%
%

\end{abstract}

\vspace{1mm}

%
\IEEEpeerreviewmaketitle

\glsresetall

\section{Introduction}
As wireless communication technologies permeate a wide range of new applications~\cite{Chen2022, Maddikunta2022,Savic2022}, the significance of location data in modern systems has become equivalent to that of communication payload data.
Applications of the \ac{IoT} frequently depend on networks consisting of numerous sensor terminals (\textit{nodes}).
Consequently, low-complexity localization algorithms that can simultaneously estimate the positions of multiple targets using aggregated multidimensional information from the nodes are highly important.
 
%
%
%
%

Within this context, we focus on a low-complexity algorithm that utilizes the isometric embedding technique, commonly referred to as \ac{MDS} \cite{Torgerson1952}, which possesses a clearly defined computational complexity, unlike methods based on Bayesian inference~\cite{Xiong2024} and convex optimization~\cite{Shi2017}.
%
%
%
%
The most widely used \acs{MDS}-based localization framework is \ac{SMDS}~\cite{Abreu2007,Macagnano2011}, which can simultaneously process hybrid information (\textit{i.e.}, both distances and angles) and has been shown to significantly outperform the classical \acs{MDS} even in the presence of angular uncertainties of approximately $\pm 35^\circ$ \cite{Macagnano2013}.

\Ac{CD-SMDS} is a technique that decreases complexity and enhances precision by tailoring \acs{SMDS} for \ac{2D} localization~\cite{Ghods2018}.
Unlike the conventional \acs{SMDS}, which uses real-valued vectors of two dimensions to describe the \ac{2D} coordinates of nodes, this approach uses complex-valued scalars to represent them and recasts the \acs{SMDS} algorithm onto the complex domain.
Constructing the \ac{GEK} matrix in the complex domain, which consolidates all measurement data, allows for a rank reduction to 1 and enhances accuracy by optimizing noise reduction via low-rank truncation using \ac{SVD}~\cite{Nishi2023}.

%
%
%
%
%

Motivated by these studies, in this paper, we propose a novel quaternion-domain SMDS (\acs{QD-SMDS}) algorithm that focuses on \ac{3D} localization by reformulating the \acs{SMDS} algorithm within the quaternion domain.
Quaternions are an extension of complex numbers, and are composed of one real part and three imaginary parts, providing four degrees of freedom~\cite{Baek2017}.
By representing \ac{3D} coordinates using three of these degrees of freedom and formulating a \ac{GEK} matrix in the quaternion domain, it is feasible to reduce the rank to $1$.
This should enhance localization precision by optimizing the noise reduction impact.

%
%
%

\textit{Notation}: Sets of real and quaternion numbers are denoted by $\mathbb{R}$ and $\mathbb{H}$, respectively.
Vectors and matrices are denoted by lower- and upper-case bold-face letters, respectively.
%
The conjugate and transpose operators are denoted by $(\cdot)^*$ and $(\cdot)^\mathsf{T}$, respectively.
In the quaternion notation, imaginary units are denoted by $\mathbf{i}$, $\mathbf{j}$, and $\mathbf{k}$, respectively, and they satisfy the relationship: $\mathbf{i}^2=\mathbf{j}^2=\mathbf{k}^2=\mathbf{i}\mathbf{j}\mathbf{k}=-1$. 
%
%
The $a\times a$ square identity matrix is denoted by $\bm{I}_{a}$.
The $a\times b$ all-zeros matrix and all-ones matrix are denoted by $\bm{0}_{a\times b}$ and $\bm{1}_{a\times b}$, respectively.
%
%
The Euclidean norm and Frobenius norm are denoted by $\|\cdot\|$ and $\|\cdot\|_{\mathrm{F}}$, respectively.
The inner and outer products are denoted by $\left<\cdot,\cdot\right>$ and $\left|\cdot\times\cdot\right|$, respectively.
%
%
%

\section{System Model}
\label{Chap:SMDS}
Consider a network embedded in a \ac{3D} Euclidean space containing $N$ nodes, out of which $N_{\mathrm{A}}$ nodes are referred to as \acp{AN}, whose locations are known without errors, while the locations of the remaining $N_{\mathrm{T}} \triangleq N - N_{\mathrm{A}}$ nodes, hereafter referred to as \acp{TN}, are to be estimated.
It is assumed that relative distances and angles between any pair of \acp{AN} or \ac{AN} and \ac{TN} are measurable, but those among \acp{TN} are not.
The goal of this paper is then to estimate the coordinates of \acp{TN} based on the measured values (including errors) between \acp{AN} and \acp{TN}, as well as the known coordinates of \acp{AN}.

Let the coordinates of the $n$-th node in the network be represented by the column vector $\bm{x}_n \triangleq \left[a_n \; b_n \; c_n\right]^\mathsf{T}\in\mathbb{R}^{3\times 1}$, which simply represents the \ac{3D} coordinates of the node in the Cartesian coordinate system.
Defining the coordinate matrix that arranges the coordinate vectors of \acp{AN} as $\bm{X}_{\mathrm{A}} \triangleq \left[\bm{x}_1\ldots,\bm{x}_{N_{\mathrm{A}}}\right]^{\mathsf{T}}\in\mathbb{R}^{N_{\mathrm{A}}\times 3}$ and coordinate matrix that arranges the coordinate vectors of \acp{TN} as $ \bm{X}_{\mathrm{T}}
\triangleq\left[\bm{x}_1,\ldots,\bm{x}_{N_{\mathrm{T}}}\right]^{\mathsf{T}}\in\mathbb{R}^{N_{\mathrm{T}}\times 3}$, the real-valued matrix that arranges the coordinate vectors of all nodes in the network can be expressed as
\begin{equation}
\label{eq:r_X}
    \bm{X}
    \triangleq
    \left[
    \bm{x}_1,\ldots,\bm{x}_n,\ldots,\bm{x}_{N}
    \right]^{\mathsf{T}}
    =
    \left[ 
    \bm{X}_\mathrm{A}^\mathsf{T},
    \bm{X}_\mathrm{T}^\mathsf{T}
    \right]^{\mathsf{T}}
    \in
    \mathbb{R}^{N\times 3}.
\end{equation}

Consider the set of unique index pairs whose distances are measurable, \textit{i.e.}, any pair among \acp{AN} or any pair between \acp{AN} and \acp{TN}, in an ascending order: $\mathcal{M}\triangleq \left\{(1,2),\cdots,(1,N),(2,3),\cdots,(2,N),\cdots,(N_{\mathrm{A}},N)\right\}$, such that each pair $m\in\mathcal{M}$ corresponds to an edge vector $\bm{v}_m$ in the form of
\begin{equation}
\label{eq:r_v}
    \vspace{-1ex}
    \bm{v}_m \triangleq \bm{x}_i - \bm{x}_j,\quad i<j,
\end{equation}
where $d_m\triangleq \| \bm{v}_m \|$ denotes the Euclidean distance between the two nodes, \textit{i.e.}, $\bm{x}_i$ and $\bm{x}_j$.
Due to space limitations, we will omit the details, but the conventional real-domain \acs{SMDS} operates based on the real-valued system in \eqref{eq:r_X} and \eqref{eq:r_v}~\cite{Abreu2007}.
%

%

\section{QD-SMDS Algorithm}
\label{Chap:QD-SMDS}
\subsection{Derivation of the Algorithm}

In this subsection, we will describe the \ac{QD-SMDS} algorithm after casting the real-valued model onto the quaternion domain.

First, the coordinate vector $\bm{x}_{n}\in\mathbb{R}^{3\times 1}$ of a generic node $n$ in the network can be alternatively expressed by the quaternion representation $\chi_n\in\mathbb{H}$ as
\begin{equation} 
\label{eq:q_x}
\bm{x}_n =\left[a_n \; b_n \; c_n\right]^\mathsf{T}
\Longleftrightarrow
\chi_{n} = a_n+\mathbf{i}b_n+\mathbf{j}c_n+\mathbf{k}\cdot 0,
\end{equation}
where out of the four degrees of freedom, three are used for the ($\mathrm{x},\mathrm{y},\mathrm{z}$) coordinates, and the remaining one is set to $0$.
%
Accordingly, the quaternion coordinate vector corresponding to the real coordinate matrix in \eqref{eq:r_X} can be expressed as
\begin{equation}
    \bm{\chi}
    \triangleq
    \left[
     \chi_1,\ldots,\chi_N
    \right]^{\mathsf{T}}
    \in \mathbb{H}^{N\times 1}.
\end{equation}

Similarly, the edge vector $\bm{v}_m$ between any two nodes $\bm{x}_i$ and $\bm{x}_j$ in \eqref{eq:r_v} can be represented as
\begin{equation}
\label{eq:q_v}
\bm{v}_{m}
\Longleftrightarrow
\nu_m
=
\grave{a}_m
+\mathbf{i}
\grave{b}_m
+\mathbf{j}
\grave{c}_m
+\mathbf{k}
\cdot 0,
\quad
i<j,
\end{equation}
where $\grave{a}_m\triangleq a_i-a_j$, $\grave{b}_m\triangleq b_i-b_j$, and $\grave{c}_m\triangleq c_i-c_j$.

From the above, the quaternion edge vector consisting of the collection of all $M\triangleq N_{\mathrm{A}}(N_{\mathrm{A}}-1)/2+N_{\mathrm{A}}N_{\mathrm{T}}$ quaternions can be concisely written as
\begin{eqnarray}
\label{eq:q_V}
\bm{\nu}
&\triangleq&
\left[
(\chi_1-\chi_2),(\chi_1-\chi_3),\ldots,(\chi_{N_{\mathrm{A}}}-\chi_N)
\right]^\mathsf{T}\nonumber \\
&=&
\left[
\nu_1,\ldots,\nu_m,\ldots,\nu_M
\right]^{\mathsf{T}}
=
\bm{C}
\bm{\chi}
\in \mathbb{H}^{M\times 1},
\end{eqnarray}
where $\bm{C}\triangleq\left[\bm{C}_{\mathrm{AA}},\bm{C}_{\mathrm{AT}}\right]^\mathsf{T}\in\mathbb{R}^{M\times N}$ is a structure matrix based on the mutual relationship between nodes and edges. The part corresponding to edge between \acp{AN} and \acp{AN}, \textit{i.e.}, $\bm{C}_{\mathrm{AA}}\in\mathbb{R}^{N_{\mathrm{A}}(N_{\mathrm{A}}-1)/2\times N}$, can be expressed as
\begin{subequations}
\label{eq:C}
\begin{equation}
    \bm{C}_{\mathrm{AA}}
\!\triangleq\!
\left[
\begin{array}{c|c|c|c|c|c}
\!\!\!\bm{1}_{N_{\mathrm{A}}-1\times 1}\! &\multicolumn{4}{c|}{-\bm{I}_{N_{\mathrm{A}}-1}} &\!\!\bm{0}_{N_{\mathrm{A}}-1 \times N_{\mathrm{T}}} \!\!\!\!\!\\\hline
\!\!\!\bm{0}_{N_{\mathrm{A}}-2\times 1}\!&\!\!\bm{1}_{N_{\mathrm{A}}-2\times 1} \!\! & \multicolumn{3}{c|}{-\bm{I}_{N_{\mathrm{A}}-2}} & \!\!\bm{0}_{N_{\mathrm{A}}-2 \times N_{\mathrm{T}}}\!\!\!\!\! \\\hline
\multicolumn{2}{c|}{\ddots}&\!\! \ddots \!\!&\multicolumn{2}{c|}{\ddots} &\vdots \\\hline
\multicolumn{3}{c|}{\bm{0}_{1\times N_{\mathrm{A}}-2}} &\!1  \!&\!\!\!-1 \!\!\!& \bm{0}_{1\times N_{\mathrm{T}}}
\end{array}
\right],
\end{equation}
while the part corresponding to edge between \acp{AN} and \acp{TN}, \textit{i.e.},  $\bm{C}_{\mathrm{AT}}\in\mathbb{R}^{N_{\mathrm{A}}N_{\mathrm{T}}\times N}$, can be expressed as
\begin{equation}
    \bm{C}_{\mathrm{AT}}
\!\triangleq\!
\left[\!
\begin{array}{c|c|c|c|c|c|c}
\!\!\bm{1}_{N_\mathrm{T} \times 1}\!\!&\multicolumn{5}{c|}{\bm{0}_{N_{\mathrm{T}} \times N_{\mathrm{A}}-1}}&\!\!\!{-\bm{I}_{N_{\mathrm{T}}}}\!\!\!\!  \\\hline
\!\!\bm{0}_{N_\mathrm{T}\times 1} \!\!& \!\!\bm{1}_{N_\mathrm{T} \times 1} \!\!& \multicolumn{4}{c|}{\bm{0}_{N_\mathrm{T} \times N_{\mathrm{A}}-2}}&\!\!\!-\bm{I}_{N_\mathrm{T}}\!\!\!\! \\\hline
\multicolumn{2}{c|}{\ddots}  &\!\!\ddots \!\! &\multicolumn{3}{c|}{\ddots}  & \vdots \\\hline
\multicolumn{3}{c|}{\bm{0}_{N_\mathrm{T}\times N_{\mathrm{A}}-2}} &\!\!\bm{1}_{N_\mathrm{T} \times 1}\!\!&\multicolumn{2}{c|}{\!\!\bm{0}_{N_\mathrm{T} \times 1}\!\!}&\!\!\! -\bm{I}_{N_\mathrm{T}}\!\!\!\! \\\hline
\multicolumn{4}{c|}{\bm{0}_{N_{\mathrm{T}}\times N_\mathrm{A}-1}} &\multicolumn{2}{c|}{\!\!\bm{1}_{N_\mathrm{T} \times 1}\!\!} &\!\!\! -\bm{I}_{N_\mathrm{T}}
\end{array}
\!\right].
\end{equation}
\end{subequations}

\begin{figure*}[!t]
\begin{center}
    	\subfigure[Cartesian coordinate system]{
	\includegraphics[width=0.5\columnwidth,keepaspectratio=true]{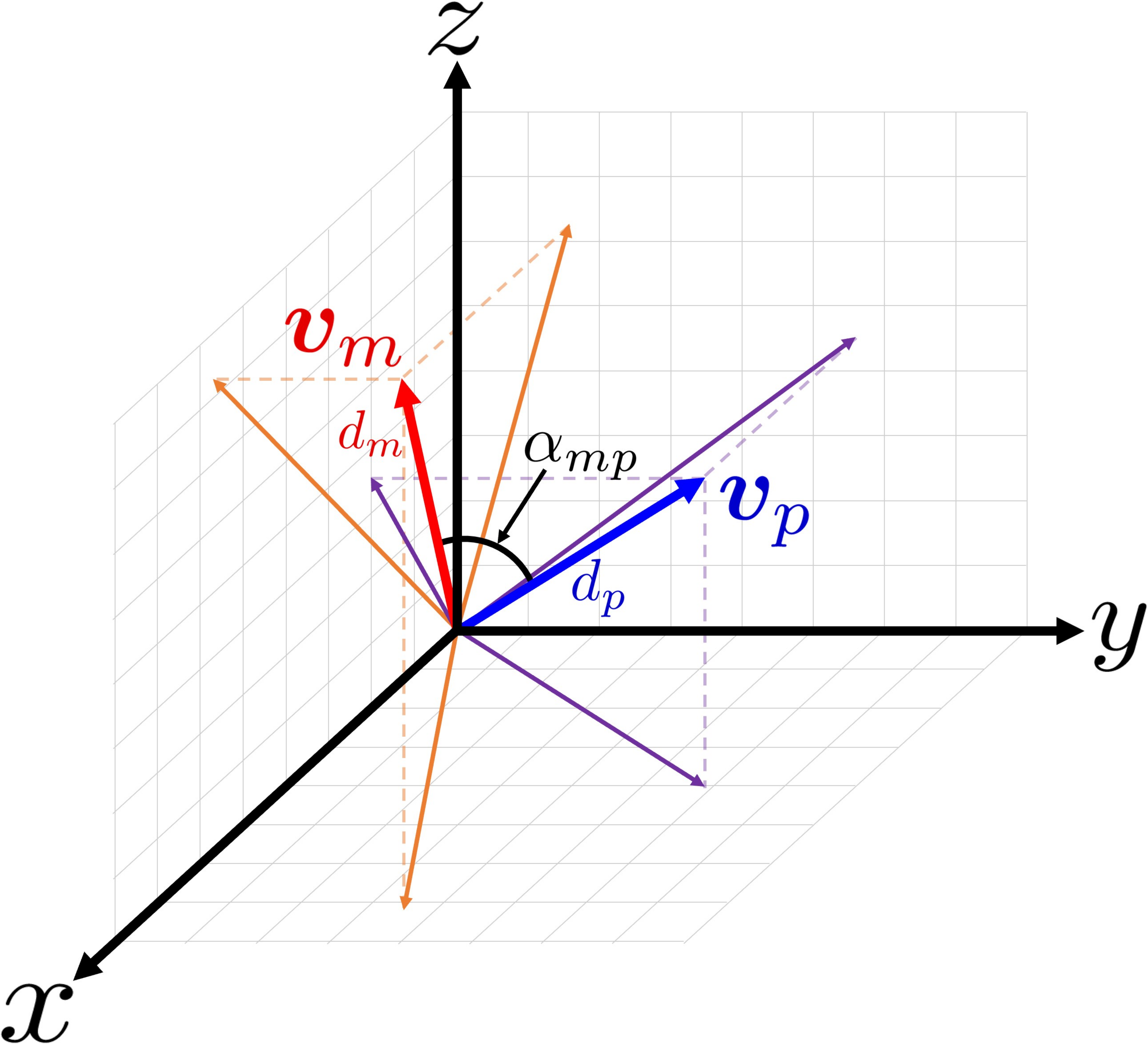}
	\label{fig:CSS}
	}
	\subfigure[$(\mathrm{x},\mathrm{y})$-plane]{
	\includegraphics[width=0.45\columnwidth,keepaspectratio=true]{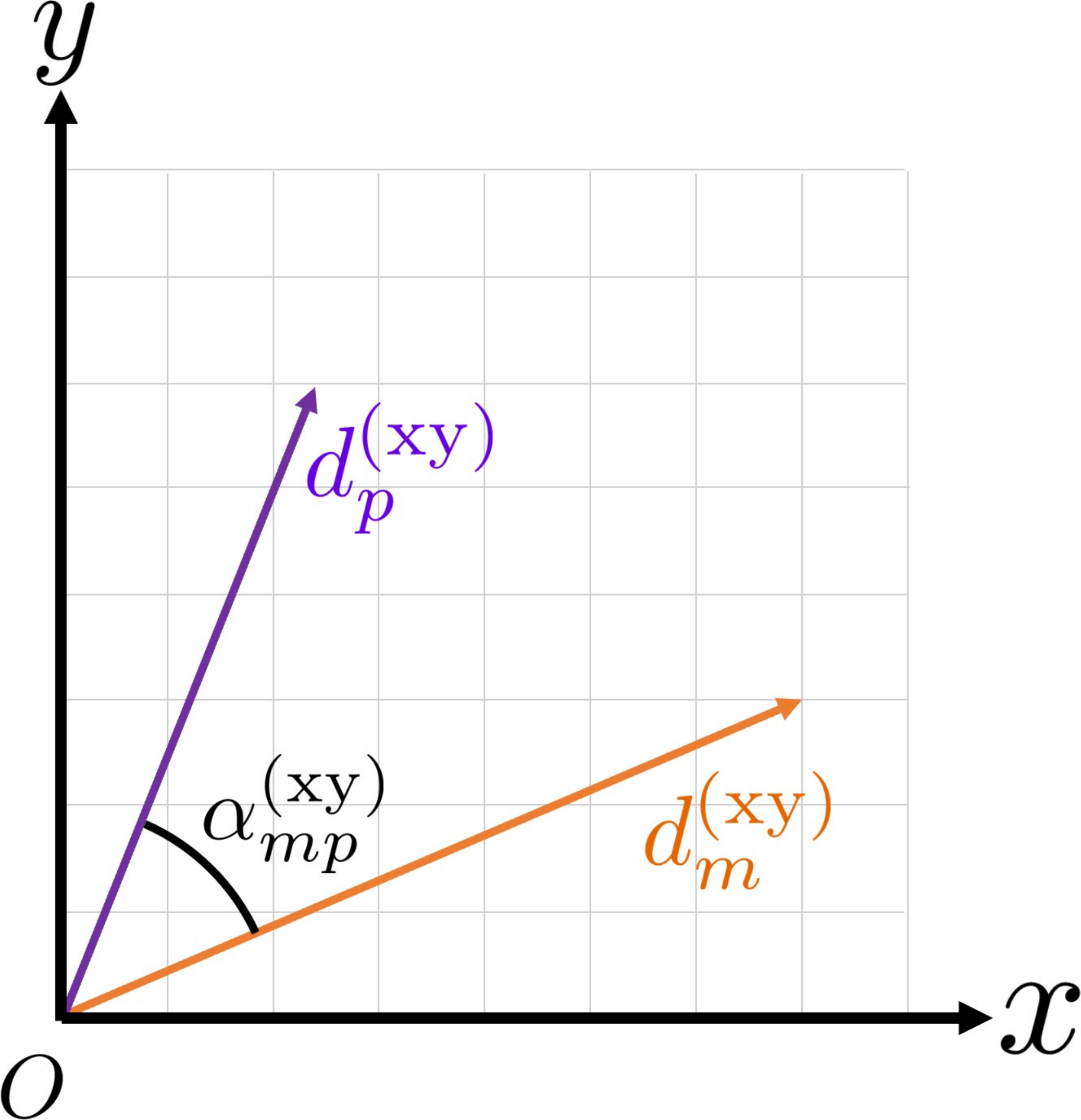}
	\label{fig:xy_plane}
	}
	\subfigure[$(\mathrm{x},\mathrm{z})$-plane]{
	\includegraphics[width=0.45\columnwidth,keepaspectratio=true]{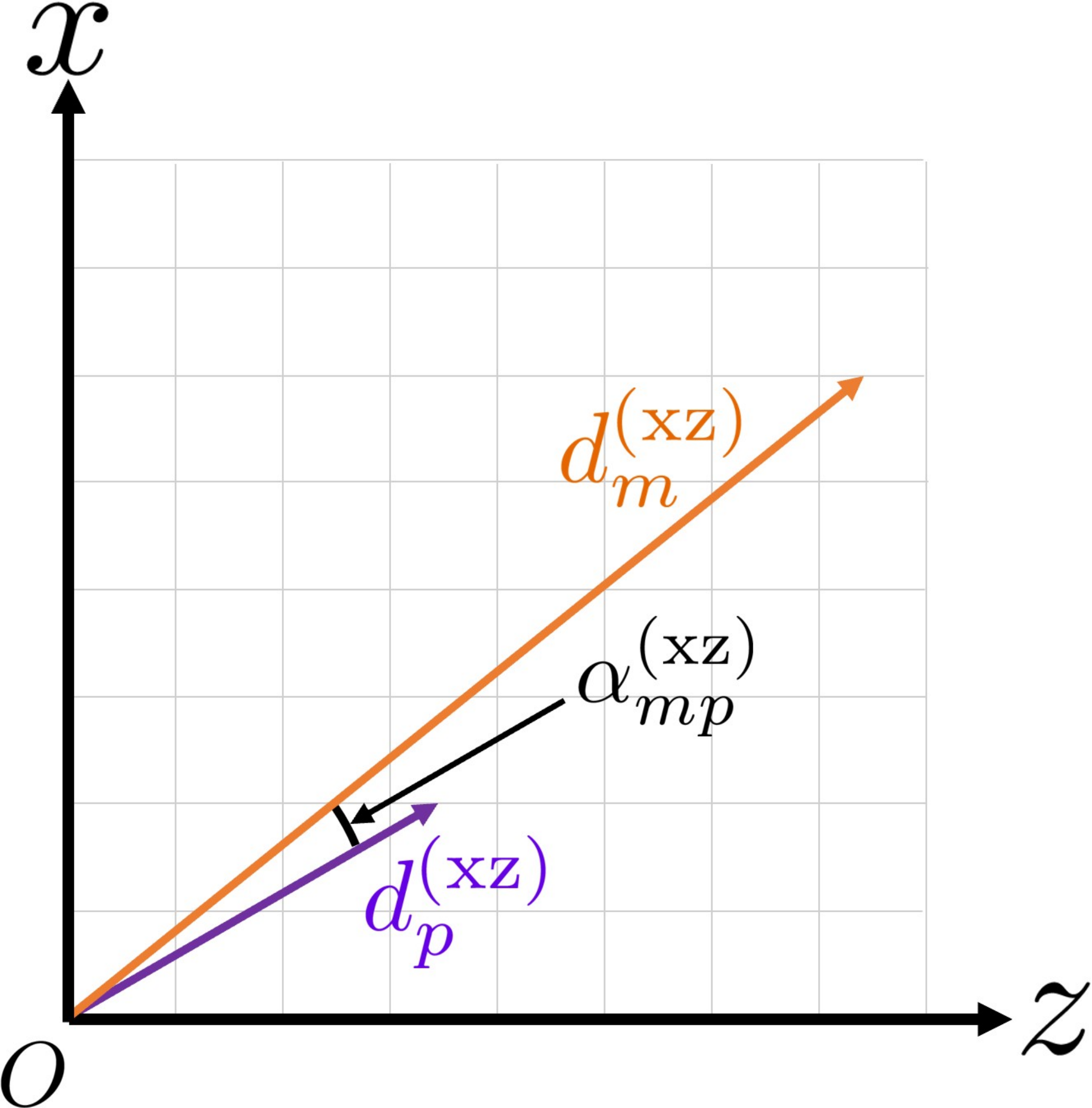}
	\label{fig:xz-plane}
	}
	\subfigure[$(\mathrm{y},\mathrm{z})$-plane]{
	\includegraphics[width=0.45\columnwidth,keepaspectratio=true]{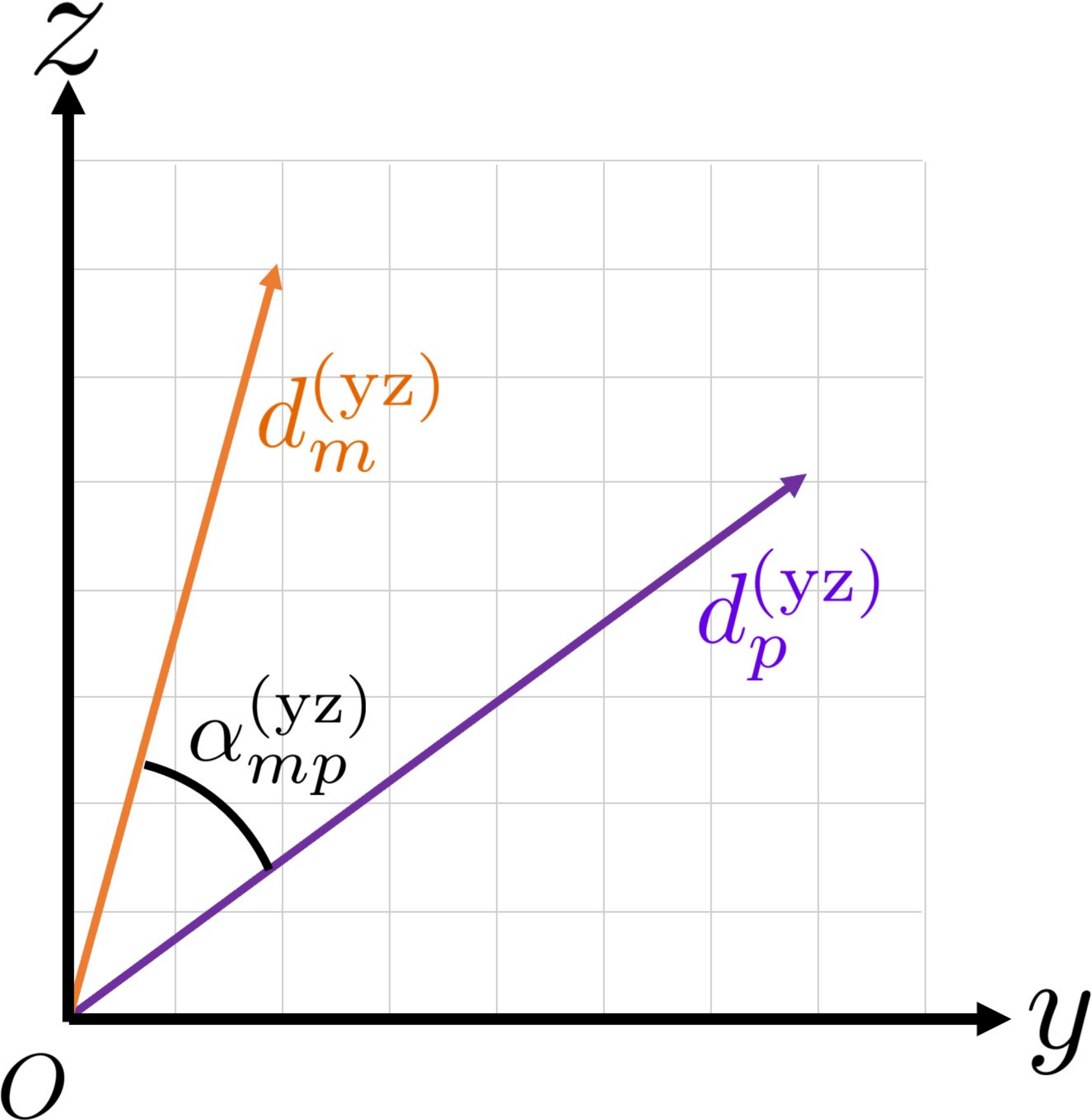}
	\label{fig:yz-plane}
	}
	\caption{
    Illustration of the parameters required to construct quaternion-domain \ac{GEK} matrix $\bm{K}$.}
    \label{fig:param_q}
	\vspace{-4ex}
\end{center}
\end{figure*}

Next, the \ac{QD-SMDS} algorithm is derived based on the quaternion system in \eqref{eq:q_x}-\eqref{eq:C}.
The inner product between the edge vectors $\bm{v}_{m}$ and $\bm{v}_{p}$ can be expressed as
\begin{eqnarray}
\left<\bm{v}_m,\bm{v}_p\right>
=
\grave{a}_m\grave{a}_p+\grave{b}_m\grave{b}_p
+\grave{c}_m\grave{c}_p
=
d_{m}d_{p}\cos\alpha_{mp},
\label{eq:inner_products}
\end{eqnarray}
where $\alpha_{mp}$ is the \ac{ADoA} between $\bm{v}_m$ and $\bm{v}_p$.

In turn, the outer product of the vectors obtained by projecting the edge vectors onto the $(\mathrm{x},\mathrm{y})$, $(\mathrm{x},\mathrm{z})$, and $(\mathrm{y},\mathrm{z})$ planes, respectively, can be expressed as
\begin{subequations}
\label{eq:outer_products}
\begin{eqnarray}
\!\!\!\!\!\!\!\!
\left|
\bm{v}_m^{(\mathrm{xy})}\!\times
\bm{v}_p^{(\mathrm{xy})}
\right|
\!\!&\!\!=\!\!&\!\!
\grave{a}_m\grave{b}_p-\grave{a}_p\grave{b}_m
=
d_m^{(\mathrm{xy})}d_p^{(\mathrm{xy})}\sin\alpha_{mp}^{(\mathrm{xy})}, \\[1ex]
\!\!\!\!\!\!\!\!
\label{eq:r_outer_yz}
\left|\bm{v}_m^{(\mathrm{xz})}\times \bm{v}_p^{(\mathrm{xz})}\right|
\!\!&\!\!=\!\!&\!\!
\grave{a}_m\grave{c}_p-\grave{a}_p\grave{c}_m
=
d_m^{(\mathrm{xz})}d_p^{(\mathrm{xz})}\sin\alpha_{mp}^{(\mathrm{xz})}, \\[1ex]
\!\!\!\!\!\!\!\!
\label{eq:r_outer_xz}
\left|
\bm{v}_m^{(\mathrm{yz})}\times
\bm{v}_p^{(\mathrm{yz})}
\right|
\!\!&\!\!=\!\!&\!\!
\grave{b}_m\grave{c}_p-\grave{b}_p\grave{c}_m
=
d_m^{(\mathrm{yz})}d_p^{(\mathrm{yz})}\sin\alpha_{mp}^{(\mathrm{yz})}, 
\end{eqnarray}
\end{subequations}
where $\bm{v}_{m}^{(\mathrm{xy})}$, $\bm{v}_m^{(\mathrm{xz})}$, and $\bm{v}_{m}^{(\mathrm{yz})}$ represent \ac{2D} vectors that are projections of $\bm{v}_m$ onto the $(\mathrm{x},\mathrm{y})$, $(\mathrm{x},\mathrm{z})$, and $(\mathrm{y},\mathrm{z})$ planes, respectively, with $d_m^{(\mathrm{xy})}\triangleq\|\bm{v}_m^{(\mathrm{xy})}\|$, $d_m^{(\mathrm{xz})}\triangleq\|\bm{v}_m^{(\mathrm{xz})}\|$, and $d_m^{(\mathrm{yz})}\triangleq\|\bm{v}_m^{(\mathrm{yz})}\|$.

Similarly, $\alpha_{mp}^{(\mathrm{xy})}$, $\alpha_{mp}^{(\mathrm{xz})}$, and $\alpha_{mp}^{(\mathrm{yz})}$ are the \acp{ADoA} between the two \ac{2D} vectors projected onto the $(\mathrm{x},\mathrm{y})$, $(\mathrm{x},\mathrm{z})$, and $(\mathrm{y},\mathrm{z})$ planes, respectively.
For a better illustration of the relationship between these parameters, please refer to Fig. \ref{fig:param_q} shown on the top of the next page.
 
Based on \eqref{eq:inner_products} and \eqref{eq:outer_products}, the product of quaternion edges $\nu_m$ and $\nu_p^*$ with $m\neq p$ can be expressed as
\begin{eqnarray}
\label{eq:q_product}
\hspace{-5ex}\nu_{m}\nu_{p}^*
\hspace{-4ex} && =
\underbrace{
\left(\grave{a}_m\grave{a}_p+\grave{b}_m\grave{b}_p+\grave{c}_m\grave{c}_p\right)
}_{\left<\bm{v}_m,\bm{v}_p\right>}
+
\mathbf{i}
\underbrace{
\left(\grave{a}_p\grave{b}_m-\grave{a}_m\grave{b}_p\right)
}_{-\left|\bm{v}_{m}^{(\mathrm{xy})}\times \bm{v}_{p}^{(\mathrm{xy})}\right|}
\nonumber\\
&&\quad + \mathbf{j}
\underbrace{
\left(\grave{a}_p\grave{c}_m-\grave{a}_m\grave{c}_p\right)
}_{-\left|\bm{v}_m^{(\mathrm{xz})}\times \bm{v}_p^{(\mathrm{xz})}\right|}
+
\mathbf{k}
\underbrace{
\left(\grave{b}_p\grave{c}_m-\grave{b}_m\grave{c}_p\right)
}_{-\left|\bm{v}_{m}^{(\mathrm{yz})}\times \bm{v}_p^{(\mathrm{yz})}\right|}
\nonumber\\
&&  = d_{m}d_{p}\cos\alpha_{mp}
-
\mathbf{i}
d_m^{(\mathrm{xy})}d_p^{(\mathrm{xy})}\sin\alpha_{mp}^{(\mathrm{xy})}\nonumber \\
&&\quad -
\mathbf{j}
d_m^{(\mathrm{xz})}d_p^{(\mathrm{xz})}\sin\alpha_{mp}^{(\mathrm{xz})} -\mathbf{k}
d_m^{(\mathrm{yz})}d_p^{(\mathrm{yz})}\sin\alpha_{mp}^{(\mathrm{yz})}.
\end{eqnarray}

Accordingly, the rank-$1$ quaternion-domain \ac{GEK} matrix that integrates all mutual distance and \ac{ADoA} information can be expressed as
\begin{equation}
\label{eq:q_kernel}
    \bm{K}
\triangleq
\bm{\nu}\bm{\nu}^\mathsf{H} 
=
\begin{bmatrix}
\nu_1\nu_1^{*} \!\!\!\!& \cdots &\!\!\!\! \nu_1\nu_M^{*}  \\
\vdots & \ddots & \vdots \\
\nu_M\nu_1^{*} \!\!\!\!& \cdots &\!\!\!\! \nu_M\nu_M^{*}
\end{bmatrix}.
\end{equation}
%
%
%
\begin{algorithm}[H]
\caption{Quaternion-Domain SMDS}\label{alg:QD-SMDS}
\begin{algorithmic}[1]
\STATE {\bf{Input:}}
\STATE \textit{Measured and estimated mutual distances and phase differences:
$d_{m},d_m^{(\mathrm{xy})},d_m^{(\mathrm{xz})},d_m^{(\mathrm{yz})},\alpha_{mp},\alpha_{mp}^{(\mathrm{xy})},\alpha_{mp}^{(\mathrm{xz})},\alpha_{mp}^{(\mathrm{yz})}$}
\STATE \textit{The coordinates of at least $4$ ANs.}
\STATE {\bf{Steps:}}
\STATE \textit{Construct the quaternion-domain \ac{GEK} matrix $\tilde{\bm{K}}$ in \eqref{eq:q_kernel} using the input parameters.}
\STATE \textit{Perform \ac{QSVD} of the constructed \ac{GEK} matrix $\tilde{\bm{K}}$} \text{(see \cite{Miao2022})}
\STATE \textit{Obtain the quaternion edge vector $\hat{\bm{\nu}}$ using \eqref{eq:q_v_estimated}}
\STATE \textit{Convert the estimated quaternion edge vector $\hat{\bm{\nu}}$ to the estimated real-valued edge matrix $\hat{\bm{V}}$.}
\STATE \textit{Compute $\hat{\bm{X}}$ from $\hat{\bm{V}}$ using \eqref{eq:MoorePenrose}}
\STATE \textit{Apply the Procrustes transform to $\hat{\bm{X}}$ if needed} \text{(see \cite{Fiore2001}).}
\end{algorithmic}
\end{algorithm}

Assuming that measured (estimated) values of all mutual distance and \ac{ADoA} parameters appearing in \eqref{eq:q_product} are available, it is evident that the quaternion-domain \ac{GEK} matrix with errors $\tilde{\bm{K}}$ can be obtained. Therefore, the estimate of the quaternion edge vector $\bm{\nu}$ is given as
\begin{equation}
\label{eq:q_v_estimated}
    \hat{\bm{\nu}}=\sqrt{\lambda}\bm{u},
\end{equation}
where $(\lambda,\bm{u})$ is the pair of the largest singular value and the corresponding singular vector of $\tilde{\bm{K}}$.
Here, we use the \ac{QSVD} proposed in \cite{Miao2022} for the \ac{SVD} of the quaternion matrix.

In the conventional \ac{SMDS} algorithm, the \ac{GEK} matrix is constructed based on the real-valued vector in \eqref{eq:r_v}; hence, its rank is $3$, and the noise reduction effect via low-rank approximation using \ac{SVD} is limited.
Meanwhile, in the \ac{QD-SMDS} algorithm, the rank of the quaternion-domain \ac{GEK} matrix is $1$, so the noise reduction effect can be maximized.

Finally, we estimate the real-valued coordinate matrix $\bm{X}$ in \eqref{eq:r_X} from the estimated quaternion edge vector $\hat{\bm{\nu}}$.
First, we take the real part, $\bm{\mathrm{i}}$-th part, and $\bm{\mathrm{j}}$-th part of $\hat{\bm{\nu}}$, and rearrange them in accordance with the $(\mathrm{x},\mathrm{y},\mathrm{z})$ coordinates based on \eqref{eq:q_v} to obtain the estimated real-valued edge matrix as
\begin{equation}
\label{eq:V}
    \hat{\bm{V}} \triangleq \left[\hat{\bm{v}}_1 , \cdots , \hat{\bm{v}}_M \right]^{\mathsf{T}} \in \mathbb{R}^{M \times 3}.
\end{equation}
Next, using the relationship between the edge and node coordinates, \textit{i.e.}, $\bm{C}$ in \eqref{eq:C}, we can calculate an estimate $\hat{\bm{X}}$.
However, since the rank of $C$ is $N-1$, the knowledge of the coordinate matrix $\bm{X}_{\mathrm{A}}$ corresponding to \acp{AN} is exploited to circumvent the rank-deficient problem.
%
%
Consequently, the estimate coordinate matrix can be recovered from $\hat{\bm{V}}$ in \eqref{eq:V} by inverting the relationship in \eqref{eq:q_V}, \textit{i.e.},
%
\vspace{-0.5ex}
\begin{equation}
\label{eq:MoorePenrose}
\left[
\begin{array}{c}
  \bm{X}_{\mathrm{A}}\\\hline
  \hat{\bm{X}}
\end{array}
\right]
=
\left[
\begin{array}{c|c}
  \bm{I}_{N_{\mathrm{A}}} & \bm{0}_{N_{\mathrm{A}}\times N_{\mathrm{T}}}\\\hline
  \multicolumn{2}{c}{\bm{C}}
\end{array}
\right]^{-1}
\left[
\begin{array}{c}
  \bm{X}_{\mathrm{A}}\\\hline
  \hat{\bm{V}}
\end{array}
\right].
\end{equation}

Finally, since the \ac{SMDS} algorithm is constructed using only the relative relationships among nodes, the inverse problem in \eqref{eq:MoorePenrose} can be characterized in various manners, and there exists a multitude of solutions.
Hence, a Procrustes transformation~\cite{Fiore2001} may be required to bring the resulting estimate $\hat{\bm{X}}$ to the same scale, orientation, and coordinate of the true coordinates $\bm{X}$. 

For the sake of completeness, we summarize the \ac{QD-SMDS} scheme in the form of a pseudo-code in Algorithm \ref{alg:QD-SMDS}.
\vspace{-1ex}
\subsection{Construction of Quaternion-Domain \ac{GEK} Matrix}
In order to construct the quaternion-domain \ac{GEK} matrix in \eqref{eq:q_kernel}, in addition to the normally measurable mutual distance $d_m$ and \ac{ADoA} $\alpha_{mp}$, additional phase difference information is required when the position relationship between nodes is projected onto every plane.
Depending on the extent to which this additional information can be obtained as measurement values, two practical scenarios are considered.
\vspace{-1ex}
%
\begin{figure}[H]
\begin{center}
	\includegraphics[width=0.8\columnwidth,keepaspectratio=true]{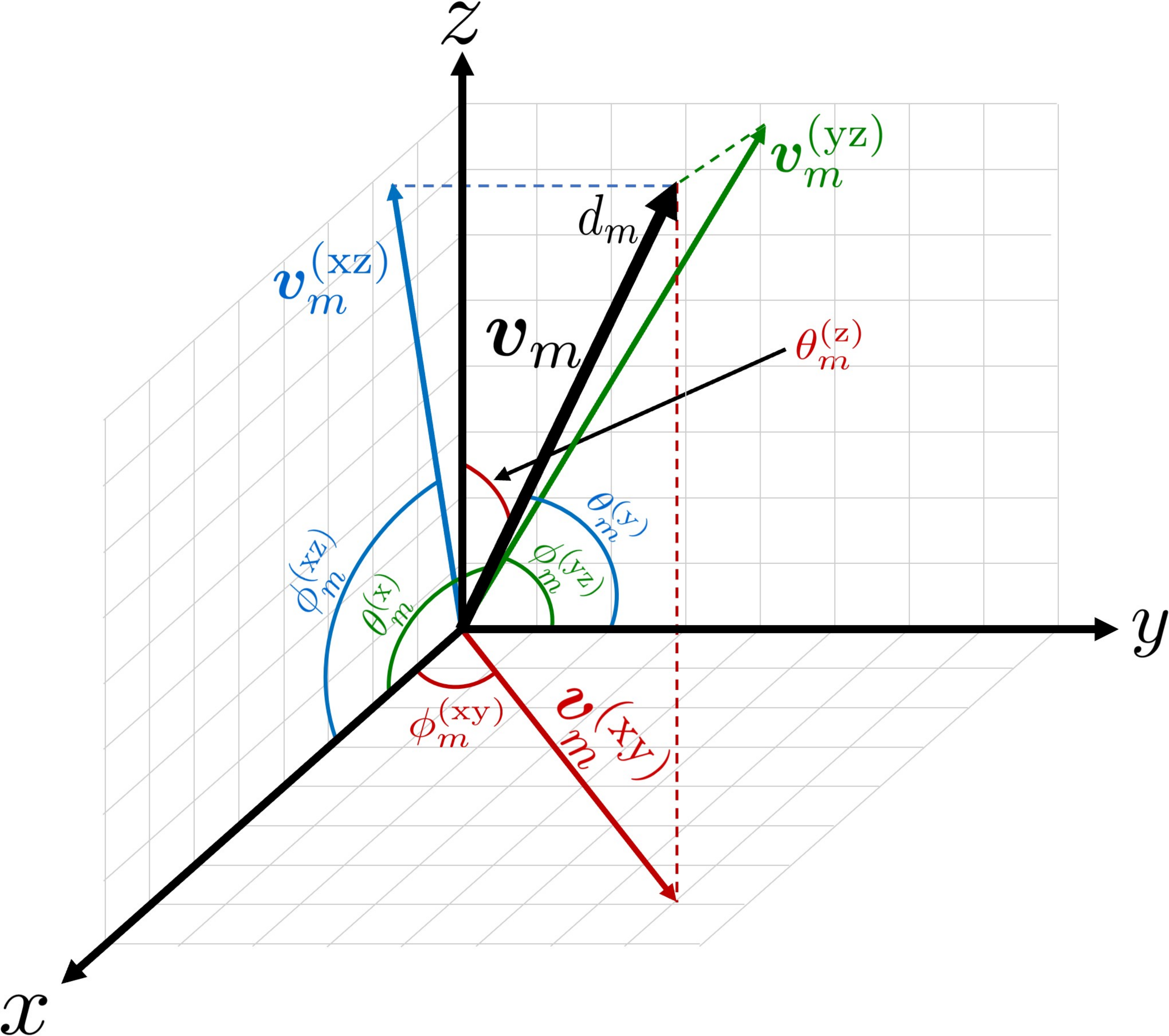}
	\caption{Parameters that can be obtained using a planar antenna.}
	\label{fig:azimuth,elevation}
	\vspace{-2ex}
\end{center}
\end{figure}

The first scenario, \textbf{Scenario II}, is one in which only the mutual distance $d_m$ and \ac{ADoA} $\alpha_{mp}$ between nodes are available, with no additional angular information obtained.
In this scenario, the quaternion-domain \ac{GEK} matrix cannot be constructed. 
Therefore, it is necessary to first execute the conventional \ac{SMDS} algorithm~\cite{Abreu2007}, calculate the angle information necessary to construct the quaternion-domain \ac{GEK} matrix from the estimated coordinates of \acp{TN} obtained, and then execute the \ac{QD-SMDS} algorithm.

%
The other scenario, \textbf{Scenario II}, is one in which the azimuth and elevation angles can be measured by using a planar antenna as proposed in~\cite{Zhang2018} at each \ac{AN}.
By appropriately arranging the planar antenna, it is possible to measure and calculate the parameters shown in Fig. \ref{fig:azimuth,elevation}, where $\theta$ represents the elevation angle, and $\phi$ represents the azimuth angle. 
%
In this scenario, we can directly execute the \ac{QD-SMDS} algorithm.

%
%

%

%

\section{Performance Assessment}
\label{Chap:simu}
\subsection{Simulation Conditions}
Computer simulations were conducted to validate the performance of the proposed \ac{QD-SMDS} algorithm.
The simulation environment is assumed to be a room with dimensions of $30\text{[m] (length)}\times30\text{[m] (width)}\times10\text{[m] (height)}$. 
The \acp{AN} were placed at five locations: the four upper corners of the room, specifically at $(x,y,z)\!\!=\!\!(0,0,10)$, $(30,0,10)$, $(30,30,10)$, and $(0,30,10)$, as well as the origin $(x,y,z)=(0,0,0)$. The \acp{TN} were randomly placed at $15$ locations in its interior with $\mathrm{x}$, $\mathrm{y}$, and $\mathrm{z}$ coordinates following a uniform distribution.

Distance measurements are modeled as Gamma-distributed random variables~\cite{Papoulis2002} with the mean given by the true distance $d$ and a standard deviation $\sigma_d$.
The \ac{PDF} of measured distances $\tilde{d}$ associated with $d$ is given by
\begin{equation}
\label{eq:pdf_d}
    p_{\mathrm{D}}(d;\alpha.\beta)
    =
    \left(\beta^\alpha
    \Gamma(\alpha)
    \right)^{-1}
    \tilde{d}^{\left(\alpha-1\right)}
    e^{-\frac{\tilde{d}}{\beta}},
\end{equation}
where $\alpha\triangleq d^2/\sigma^2_d$ and $\beta\triangleq \sigma^2_d/d$.

In turn, angle measurement errors $\delta_{\theta}$ are assumed to follow Tikhonov-distribution~\cite{Abreu2008}.
The \ac{PDF} of measured angles $\tilde{\theta} = \theta + \delta_{\theta}$ associated with a true angle $\theta$ is given by
\begin{equation}
p_{\mathrm{\Theta}}\left(\tilde{\theta};\theta,\rho \right)
=
\frac{1}{2\pi I_0(\rho)}\exp\left[\rho\cos(\theta-\tilde{\theta})\right].
\end{equation}

The range of the angular error is influenced by the angular parameter $\epsilon$, which represents the bounding angle of central $90^{\mathrm{th}}$ percentile, expressed as
\begin{equation}
\label{eq:angle_error}
\epsilon=\theta_{\mathrm{B}} \Big| \int_{-\theta_{\mathrm{B}}}^{\theta_{\mathrm{B}}} p_{\Theta}(\phi ; 0, \rho) d\phi=0.9.
\vspace{-1ex}
\end{equation}

Estimation errors are measured by the Frobenius norm of the difference between the estimates $\hat{\bm{X}}$ and true \acp{TN}' position $\bm{X}$,
\vspace{-1ex}
\begin{equation}
\label{eq:mse}
\xi=\frac{1}{N_{t}}\|\hat{\bm{X}}-\bm{X}\|_{\mathrm{F}}. 
\end{equation}

\subsection{Simulation Results}
\begin{figure}[H]
\begin{center}
	\subfigure{
	\includegraphics[width=\columnwidth]{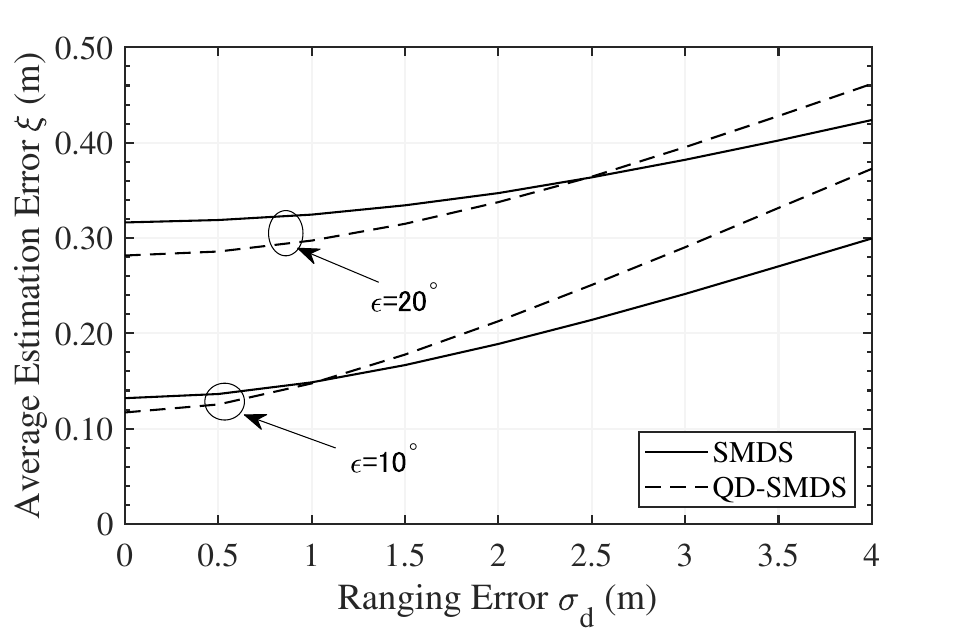}
    \label{fig:unknown10-30}
    }
	\subfigure{
	\includegraphics[width=\columnwidth]{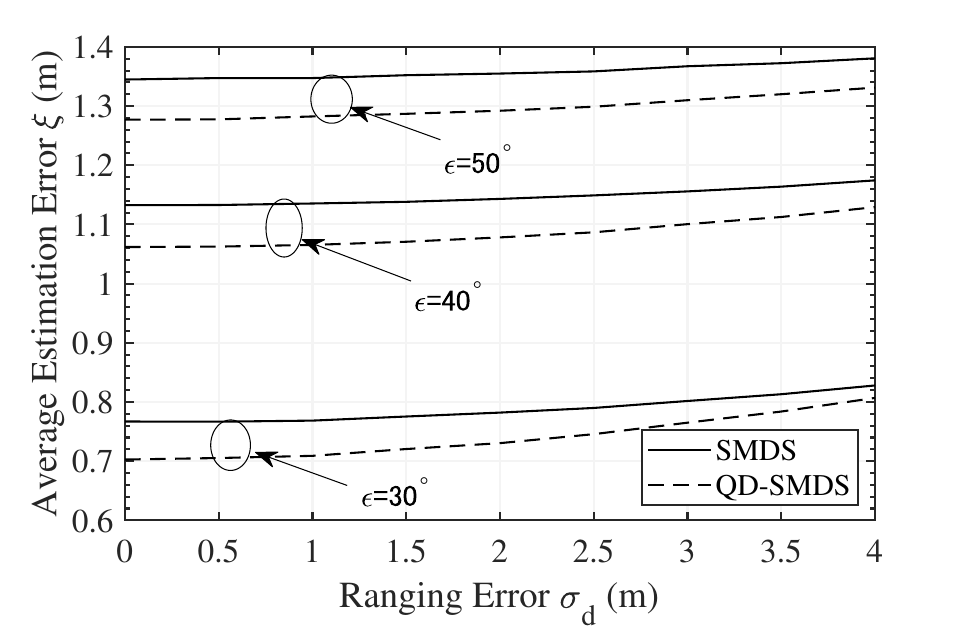}
	}
    \vspace{-4ex}
	\caption{Comparison of localization accuracy between the \ac{SMDS} and \ac{QD-SMDS} algorithms in \textbf{Scenario I}, where only mutual distance and \ac{ADoA} between nodes are available.}
	\label{fig:SMDSvsQDSMDSvol1}
	\vspace{-2ex}
\end{center}
\end{figure}

Based on the above simulation conditions, we have compared the localization accuracy of the conventional \ac{SMDS} and proposed \ac{QD-SMDS} algorithms for each of \textbf{Scenario I} and \textbf{Scenario II} described in the previous section.
%
%
Fig. \ref{fig:SMDSvsQDSMDSvol1} shows a comparison of localization accuracy between the \ac{SMDS} and \ac{QD-SMDS} algorithms in \textbf{Scenario I}. 
The horizontal axis denotes the standard deviation of Gamma-distributed distance measurements in \eqref{eq:pdf_d}, while the vertical axis denotes the averaged estimation error in \eqref{eq:mse}.
The localization accuracy was plotted with different angular measurement errors $\epsilon\in\left\{ 10^\circ,20^\circ,30^\circ,40^\circ,50^\circ \right\}$.

When the angle error is small ($\epsilon = 10^\circ$ and $20^\circ$), we can observe that the superiority of the two methods switches depending on the distance error.
The \ac{QD-SMDS} method performs better than the \ac{SMDS} algorithm up to $\sigma_d = 1.0$ [m] for $\epsilon=10^\circ$, and up to $\sigma_d = 1.8$ [m] for $\epsilon=20^\circ$; however, when the error increases, the SMDS algorithm performs better.
%
%
The reason for this is that, when the angle error is minimal, the \ac{GEK} matrix can be constructed with a high degree of accuracy.
Consequently, maximizing noise reduction through low-rank approximation is not critically important.
Additionally, the \ac{SMDS} algorithm, which is capable of constructing the \ac{GEK} matrix with fewer (noisy) parameters, can function with enhanced accuracy.

\newpage

%
\vspace{-2ex}
\begin{figure}[H]
\begin{center}
	\subfigure{
	\includegraphics[width=\columnwidth]{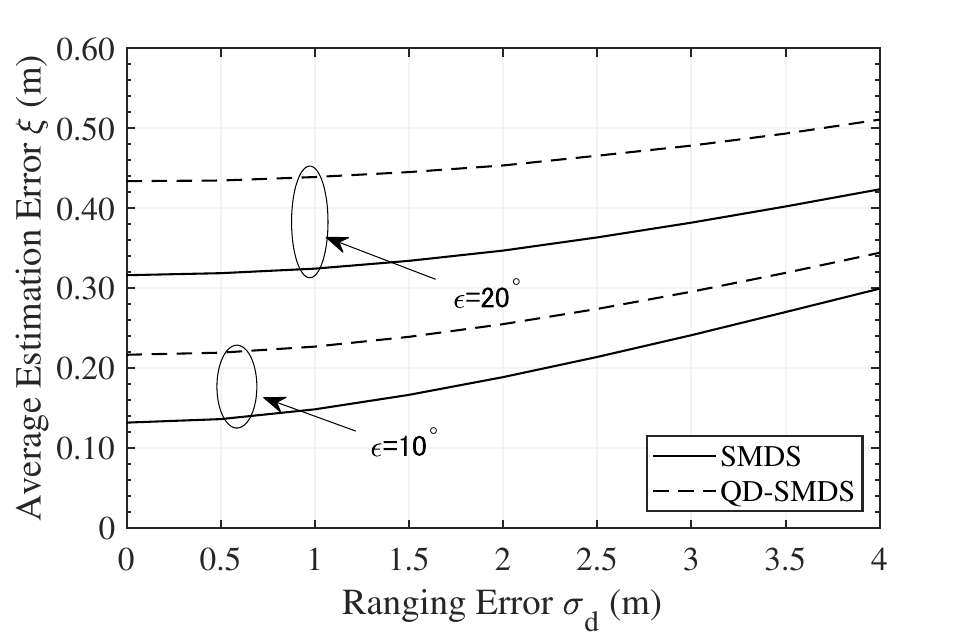}
	}
	\subfigure{
	\includegraphics[width=\columnwidth]{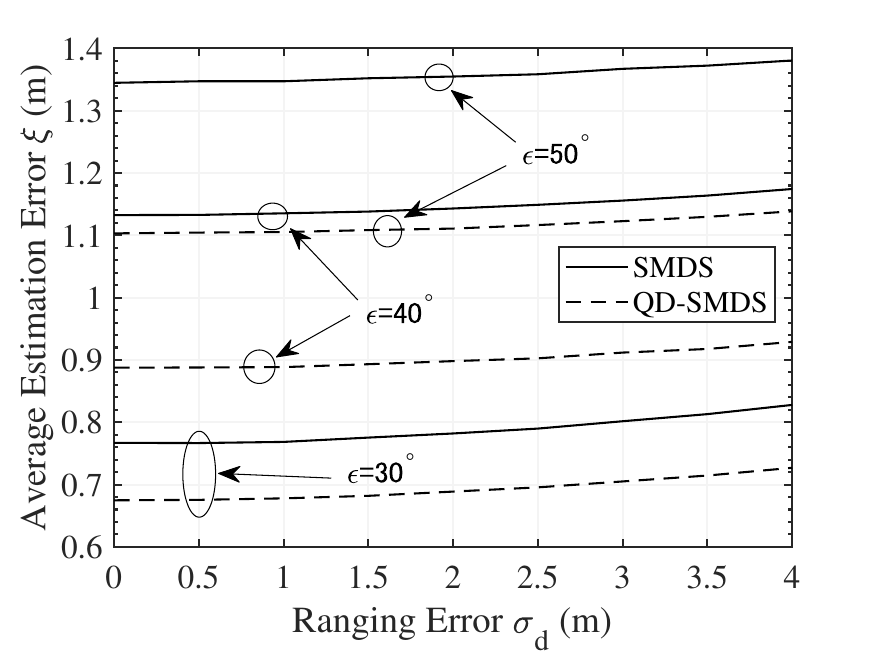}
	}
    \vspace{-2ex}
	\caption{Comparison of localization accuracy between the \ac{SMDS} and \ac{QD-SMDS} algorithms in \textbf{Scenario II}, where all data constituting the quaternion-domain \ac{GEK} matrix in \eqref{eq:q_kernel} is accessible.}
	\label{fig:SMDSvsQDSMDSvol2}
    \vspace{-2ex}
\end{center}
\end{figure}

In contrast, when the angle error is $30^\circ$ or more, the \ac{QD-SMDS} algorithm always achieves better performance than the \ac{SMDS} algorithm, and the performance difference becomes more pronounced as the angle error increases.
%

The results indicate that as the accuracy of the \ac{GEK} matrix decreases, the significance of the noise reduction effect achieved through low-rank approximation using \ac{SVD} increases.
%
%
Therefore, the \ac{QD-SMDS} algorithm with a rank-$1$ \ac{GEK} matrix demonstrates enhanced robustness in the presence of angle errors compared to the \ac{SMDS} algorithm employing a rank-$3$ \ac{GEK} matrix.

Fig. \ref{fig:SMDSvsQDSMDSvol2} shows a comparison of localization accuracy between the \ac{SMDS} and \ac{QD-SMDS} algorithms in \textbf{Scenario II}. 
As with the results in Fig. \ref{fig:SMDSvsQDSMDSvol1}, \ac{SMDS} achieves better localization accuracy up to $\epsilon = 20^\circ$. 
In contrast, when the angle error increases, the \ac{QD-SMDS} algorithm significantly outperforms the \ac{SMDS} algorithm, and the performance difference is greater than that in \textbf{Scenario I} shown in Fig. \ref{fig:SMDSvsQDSMDSvol1}. 
This is due to the fact that the robustness of the \ac{QD-SMDS} algorithm against errors has improved by the additional angle information that can be measured and estimated using a planar antenna.

%
Based on the numerical results and the computational effort involved, it is determined that the \ac{SMDS} algorithm is preferable for scenarios with relatively small angle errors, while the \ac{QD-SMDS} is more suitable for cases with larger angle errors. 
Furthermore, it was shown that the supplementary data regarding azimuth and elevation angles obtained from the planar antenna array becomes increasingly critical as the angle error escalates.
%
%
%
%
%

\section{Conclusion}
\label{Chap:conc}
In this paper, we proposed a novel \ac{QD-SMDS} algorithm, which is derived by recasting the classical \ac{SMDS} algorithm onto the quaternion domain, with the aim of achieving low-complexity simultaneous localization for multiple targets using information aggregated from a large number of wireless sensor nodes.
By constructing a \ac{GEK} matrix in the quaternion domain, it is possible to reduce the rank to $1$ even for networks in \ac{3D} Euclidean space, and the noise reduction effect can be maximized via \ac{QSVD}. 
Simulation results show that the proposed method can achieve higher localization accuracy than the conventional method when the angle error is significant.



\section*{Acknowledgement}
This work was supported in part by JSPS KAKENHI Grant Number JP23K13335 and in part by JST, CRONOS, Japan Grant Number JPMJCS24N1.


\vspace{-1mm}
\begin{thebibliography}{99}

\bibitem{Chen2022}
H. Chen et al., ``A tutorial on terahertz-band localization for 6G communication systems,'' \textit{IEEE Commun. Survey Tut.}, vol. 24, no. 3, pp. 1780--1815, 2022.

\bibitem{Maddikunta2022}
P. K. R. Maddikunta et al., ``Industry 5.0: A survey on enabling technologies and potential applications,'' \textit{J. Ind. Inf. Integr.}, vol. 26, p. 100257, 2022.

\bibitem{Savic2022}
T. Savić, X. Vilajosana, and T. Watteyne, ``Constrained localization: A survey,'' \textit{IEEE Access}, vol. 10, pp. 49\,297--49\,321, 2022.

\bibitem{Torgerson1952}
W. S. Torgerson, ``Multidimensional scaling: I. theory and method,'' \textit{IEEE Trans. Signal Process.}, vol. 17, no. 4, p. 401--419, 1952.

\bibitem{Xiong2024}
J. Xiong et al., ``Adaptive message passing for cooperative positioning under unknown non-Gaussian noises,'' \textit{IEEE Trans. Instrum. Meas.}, vol. 73, pp. 1--14, 2024.

\bibitem{Shi2017}
X. Shi et al., ``Robust localization using range measurements with unknown and bounded errors,'' \textit{IEEE Trans. Wireless Commun.}, vol. 16, no. 6, pp. 4065--4078, 2017.

\bibitem{Abreu2007}
G. Abreu and G. Destino, ``Super MDS: Source location from distance and angle information,'' in Proc. \textit{IEEE Wireless Commun. Netw. Conf. (WCNC)}, vol. 2, Mar. 2007, pp. 4430--4434.

\bibitem{Macagnano2011}
D. Macagnano and G. Abreu, ``Super MDS with heterogeneous information,'' in Proc. \textit{Conf. Record of the Forty Fifth Asilomar Conf. Signals, Systems and Computers (Asilomar)}, 2011, pp. 1519--1523.

\bibitem{Macagnano2013}
D. Macagnano and G. Abreu, ``Algebraic approach for robust localization with heterogeneous information,'' \textit{IEEE Trans. Wireless Commun.}, vol. 12, no. 10, pp. 5334--5345, 2013.

\bibitem{Ghods2018}
A. Ghods and G. Abreu, ``Complex-domain super MDS: A new framework for wireless localization with hybrid information,'' \textit{IEEE Trans. Wireless Commun.}, vol. 17, no. 11, pp. 7364--7378, 2018.

\bibitem{Nishi2023}
Y. Nishi et al., ``Wireless location tracking via complex-domain super MDS with time series self-localization information,'' in Proc. \textit{2023 IEEE Intl. Conf. Acoustics, Speech and Signal Process. (ICASSP)}, 2023, pp. 1--5.

\bibitem{Baek2017}
G. K. J. Baek, H. Jeon and S. Han, ``Visualizing quaternion multiplication,'' \textit{IEEE Access}, vol. 5, pp. 8948--8955, 2017.

\bibitem{Miao2022}
J. Miao and K. I. Kou, ``Color image recovery using low-rank quaternion matrix completion algorithm,'' \textit{IEEE Trans. Image Process.}, vol. 31, pp. 190--201, 2022.

\bibitem{Fiore2001}
P. Fiore, ``Efficient linear solution of exterior orientation,'' \textit{IEEE Commun. Surveys Tuts}, vol. 23, pp. 140--148, Feb. 2001.

\bibitem{Zhang2018}
D. Zhang et al., ``Two-dimensional direction of arrival estimation for coprime planar arrays via polynomial root finding technique,'' \textit{IEEE Access}, vol. 6, pp. 19\,540--19\,549, 2018.

\bibitem{Papoulis2002}
A. Papoulis and P. S. U. Pillai, \textit{Random Variables and Stochastic Processes}, 4th ed. New York, NY, USA: McGraw-Hill, 2002.

\bibitem{Abreu2008}
G. Abreu, ``On the generation of Tikhonov variates,'' \textit{IEEE Trans. Commun.}, vol. 56, p. 1157--1168, Jul. 2008.

\end{thebibliography}
\end{document}